\documentclass[11pt,twoside]{article}

\usepackage{asp2006_hotcool}
\usepackage{epsf}
\usepackage{graphicx}
\usepackage{natbib}
\usepackage{lscape}

\def\ga{\mathrel{\hbox{\rlap{\hbox{\lower4pt\hbox{$\sim$}}}\hbox{$>$}}}}
\def\la{\mathrel{\hbox{\rlap{\hbox{\lower4pt\hbox{$\sim$}}}\hbox{$<$}}}}

\def\msun{$M$\mbox{$_{\normalsize\odot}$}}

\def\minit{$M_{\rm init}$}

\def\kms{\,km~s$^{-1}$}

\begin{document}
\title{Young Massive Clusters as probes of stellar evolution}   %%% Fill in title
\author{Ben Davies}   %%% Fill in author names
\affil{School of Physics \& Astronomy, University of Leeds,
  Woodhouse Lane, Leeds LS2 9JT, UK; \\
Chester F.\ Carlson Center for Imaging Science, Rochester
Institute of Technology, 54 Lomb Memorial Drive, Rochester NY, 14623,
USA}    %%% Fill in author affiliations

\begin{abstract} %%% Abstract to run on from here.
Young Massive Clusters (YMCs) represent ideal testbeds in which to
study massive stellar evolution as they contain large, coeval,
chemically homogeneous, samples of massive stars. By studying YMCs
with a range of ages (and hence turn-off masses), we can investigate
the post main-sequence evolution of massive stars as a function of
initial mass. Recent discoveries of YMCs over a range of ages within
our own Galaxy - where we can successfully resolve individual stars -
offers the unprecedented opportunity to test our ideas of massive
stellar evolution. Here, I review some of the recent works in this
field, and describe how we can use YMCs to investigate several topics,
including (a) the evolutionary state of H-rich Wolf-Rayet stars; (b)
the influence of binarity on stellar evolution in dense clusters; and
(c) Red Supergiants and the post-supernova remnants they leave behind.
\end{abstract}

\section{Introduction}

It is increasingly apparent that all stars form in some form of
cluster. In very massive star clusters, large numbers of massive stars
can be found, and the strong winds of these stars can very quickly
evacuate the leftover natal material, halting the star-forming
process. What is left is a large collection of stars which have
practically the same age, a freeze-frame in the evolution of stars
with a range of initial masses. By assembling a large number of such
clusters with a range of ages in which we can resolve the individual
stars, we can build a movie of stellar evolution as we watch stars of
progressively lower mass evolve off the main-sequence (MS). By
measuring the point at which stars are just leaving the MS, we can
make direct links between initial stellar mass, various post-MS
evolutionary phases and even post-SN objects such as neutron stars.

The idea of using young clusters of stars to draw conclusions about
stellar evolution is not new -- there have been several pioneering
works by, for example,
\citet{S-M84,Humphreys85,Massey00,Massey01}. However, these early
works were typically confined to optically visible clusters, and hence
objects which were within about 1kpc of the Sun. Unfortunately, the
young star clusters within this volume of space are not the most
dazzling; few have masses in excess of $10^{3}$\msun. If a typical
Salpeter Initial Mass Function (IMF) is assumed, one needs clusters
which are more massive by a factor of $\sim$10 in order to have a
statistically-useful number of massive stars.

In the last 10-15 years, advances in infra-red astronomy -- both from
the point-of-view of instrumentation and of analysis techniques --
have greatly increased the detectable volume of the Galaxy, it is now
commonplace to observe objects at near-IR wavelengths which are
obscurred by $\sim$30mags of extinction in the optical. Consequently,
the number of Galactic young star clusters with masses in excess of
the magic figure of $10^{4}$\msun\ is growing at a rate of about 1 per
year. 

Using this sample of Young, Massive Clusters (YMCs) we can now begin
to address several topics pertaining to the evolution of massive
stars, such as the nature of Luminous Blue Variables, the progenitors
of SNe, and the influence of binarity. They also offer the opportunity
to critically test models of massive stellar evolution, which are an
integral part of the study of unresolved starburst systems in external
galaxies. 

In the following sections, we will review how topics such as these
have recently been addressed in a series of case studies of individual
clusters.

\section{Case study I. The Arches Cluster}
This cluster, located only $\sim$30pc away from the Galactic Centre,
has a mass of around $2 \times 10^{4}$\msun\ and is about 2-3Myr old
\citep{Cotera96,Figer02,Najarro04,Figer05,Martins08}. Its age means
that its most massive stars are still on -- or at least are very close
to -- the MS. These stars appear as the so-called Hydrogen- and
Nitrogen-rich Wolf-Rayet stars (WNLh). They are not what one typically
means when talking about Wolf-Rayet stars (WRs); they still have
H-rich envelopes and are core H-burning objects. However, they drive
such massive, fast winds that their spectra are filled with the
strong, broad emission lines synonymous with WRs.

\begin{figure}[t]
  \centering
  \includegraphics[width=8cm]{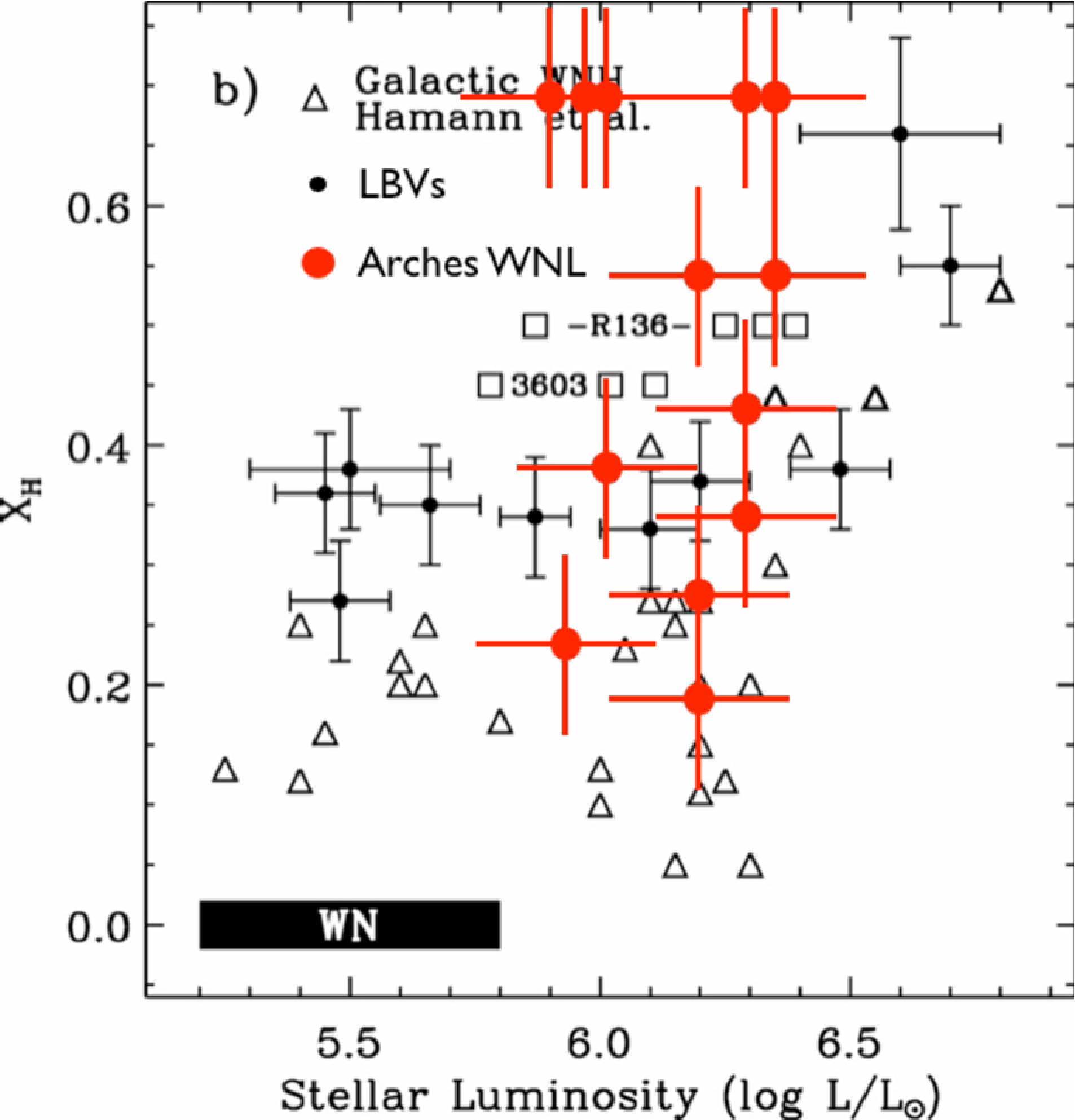}
  \caption{Hydrogen mass-fraction versus luminosity for LBVs, a
    Galactic sample of WNLh stars \citep[from][]{Hamann06} and the
    WNLh stars in the Arches Cluster. From the Galactic sample of WNLh
    stars, it appears that they are more evolved than the
    LBVs. However, the homogeneous sample of the Arches cluster
    \citep[taken from][]{Martins08} suggests that they occupy the same
    evolutionary space, and may even be less evolved than the
    LBVs. Adapted from a plot in \citet{S-C08}.}
  \label{fig:wnl}
\end{figure}

The large, homogeneous sample of WNLh stars in the Arches Cluster
offers the opportunity to re-assess the evolutionary state of these
stars with respect to other massive post-MS objects, as we can be
reasonably confident of the stars' luminosities and initial masses. In
Fig.\ \ref{fig:wnl}, we show a plot of H mass-fraction versus
luminosity for WNLh stars in comparison to Luminous Blue Variables
(LBVs), taken from \citet{S-C08}. The evolutionary path of stars on
this diagram begins at the top (i.e. H-rich), and falls downwards as
the H-fraction drops gradually to $\sim$zero, whereupon the star
becomes a `typical' WN star. By comparing the H mass fractions of LBVs
to a sample of Galactic WNLh stars from \citet{Hamann06}, Smith \&
Conti argued that the LBVs represented a phase immediately preceeding
the WNLh phase, due to the H content of LBVs being generally higher
than that of WNLh stars.

However, the \citet{Hamann06} WNLh sample was not homogeneous; it
contained objects located througout the Galaxy with presumably a large
range of initial masses, as well as uncertain distances. The Arches
Cluster sample of WNLh stars is arguably a much better tool with which
to address this problem. In Fig.\ \ref{fig:wnl} we overplot the Arches
WNLh stars, using the data from \citet{Martins08}. We can see that,
rather than being more evolved than the LBVs, many have similar H
mass-fractions. Indeed, some of the Arches WNLh stars still have all
their H.

Figure \ref{fig:wnl} would seem to cast doubt on an evolutionary
connection between WNLh stars and LBVs. Indeed, if the most massive
stars in the Arches Cluster, whose luminosities are consistent with
those of LBVs, can lose 80\% of their H without apparently going
through an LBV phase, does this suggest that not all massive stars
become LBVs? This might mean that LBVs result from stars with special
initial conditions such as fast rotation or binarity. Alternatively,
it might mean that the LBV phase is much shorter than previously
thought ($\la10^{4}$yrs), and that even in a cluster as well populated
with massive stars as The Arches the chances of catching a star as an
LBV are still small.

\section{Case study II. Westerlund~1}
At 4-5Myr old, Westerlund~1 (Wd~1) is slightly older than The
Arches. It is the most massive known young cluster in the Galaxy,
approaching $10^{5}$\msun \citep{Clark05,Brandner08}. The cluster is
notable for its rich stellar inventory -- to date, there have been 24
WRs confirmed (20\% of all known WRs in the Galaxy), as well as two
candidate LBVs, four P~Cygni-type hypergiants, many more `regular' OB
supergiants, six Yellow Hypergiants (YHGs) and four extreme
mass-losing Red Supergiants (RSGs).

\begin{figure}
  \centering
  \includegraphics[width=8cm]{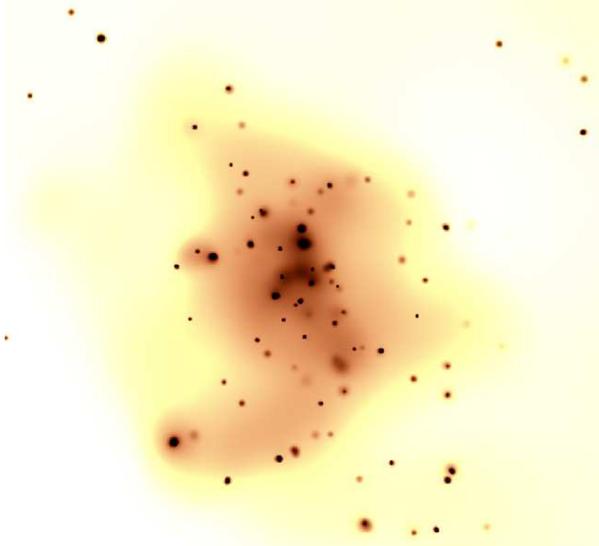}
  \caption{Westerlund~1 as seen in X-rays. Many of the large number of
  point sources can be associated with massive stars in interacting
  binary systems, which imply a lower-limit to the binary fraction of
  massive stars of $\ga$70\% \citep{Clark08}. }
  \label{fig:wd1}
\end{figure}

As Wd~1 is the closest thing we have to the so-called Super Star
Clusters observed in star-forming galaxies, the obvious challenge to
the Simple Stellar Population (SSP) models used to analyze unresolved
extra-galactic starburst systems is to reproduce the properties of
Wd~1. Perhaps predictably, they are unable to do so. Specifically, it
was shown by \citet{Crowther06} that single star evolutionary models
under-predicted the numbers of N-rich WRs with respect to H-rich and
C-rich WRs. They argued that, by incorporating the effects of close
binary evolution such as Roche-lobe overflow and common envelope
evolution, the ratios of WNh/WN and WC/WN could be reduced, and that
the observed WR population of Wd~1 was evidence for a high binary
fraction in Wd~1.

Further -- and arguably more direct -- evidence for a high binary
fraction in Wd~1 was presented in \citet{Clark08}. Using X-ray data
(see Fig.\ \ref{fig:wd1}) they showed that 17/24 WRs in Wd~1 were in
massive binary systems, whereby the wind-wind collision region between
the WR and a companion star produces a hard X-ray spectrum. That is,
70\% of stars in Wd~1 with masses greater than $\sim$40\msun\ are in
interacting binary systems. This is very much a lower limit, as X-rays
will only be produced if the separation between the two stars is large
enough to allow the winds to accelarate to $\ga$1000\kms\ before
colliding.

So, the key lesson to be learned from Wd~1 is that binarity plays a
key role in the evolution of massive stars, {\it especially} in dense
massive clusters. This point is particularly important when
interpreting the integrated properties of unresolved systems such as
young massive clusters in starbursting galaxies.

\section{Case study III. The Red Supergiant Clusters}
In recent years, two more young massive clusters have been discovered
in the inner Galaxy by virtue of the large numbers of RSGs they
contain and the brightnesses of these stars in the near-IR. The
so-called Red Supergiant Clusters (RSGCs) in the direction of Scutum
both have masses in excess of $10^{4}$\msun, with ages of 12Myr and
20Myr for clusters 1 and 2 respectively
\citep{Figer06,RSGC2paper,RSGC1paper}. Each cluster on its own
represents a large homogeneous sample of RSGs, while the two clusters
together -- with their differing ages and similar chemical abundances
\citep{RSGCabund} -- offer the opportunity to study RSG evolution at
uniform metallicity and as a function of initial mass.

Cluster \#1 (hereafter RSGC1) in particular allows us to study the
phenomenon of maser emission in RSGs. Maser emission is typically
observed in the most `extreme' RSGs such as VY~CMa and $\mu$-Cep,
objects with high mass-loss rates and large amounts of circumstellar
material. It is also seen in the proto-typical post-RSG IRC~+10420, an
object which is too hot for molecules to survive in its wind and must
therefore have evolved away from a cooler phase very quickly. So, is
the maser active phase of RSGs an indication that the star is
experiencing a very high mass-loss rate, and is perhaps on the verge of
shedding its envelope and beginning its evolution toward the blue?

\begin{figure}
  \centering
  \includegraphics[width=12cm,bb=30 20 690 500]{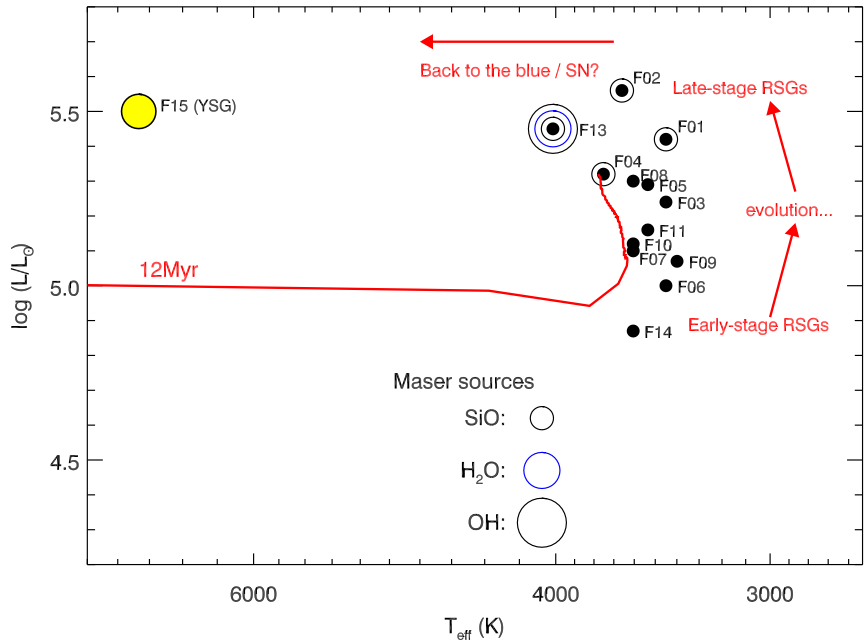}
  \caption{HR-diagram of the stars in RSGC1. Overplotted is a 12Myr
    isochrone from \citet{Mey-Mae00}, and the annotations indicate the
    evolutionary path along the isochrone. The objects with maser
    emission are indicated. }
  \label{fig:maser}
\end{figure}

We investigate this question of `masers as an evolutionary signpost'
in Fig.\ \ref{fig:maser}, where we plot a HR-diagram for the stars in
RSGC1. In the figure we identify those stars which have SiO maser
emission, which originates close to the photosphere, with a small
circle. Masers of H$_{2}$O and OH, which originate at progressively
larger distances from the star, are indicated with larger circles.

We find that only the most luminous stars in RSGC1 have maser
emission. Further, we see that only F13 -- a very bright star which
appears to be on the verge of returning toward the blue -- has all
three masers switched on. From these data, we suggest the following
scenario: once leaving the MS, stars with \minit$\sim
10-20$\msun\ evolve very quickly across the HR-diagram to become
'early-stage' RSGs. As the He core grows in mass, the star becomes
more luminous, and the increased radiation pressure on the surface
layers drives up the mass-loss rate. Eventually, the mass-loss has
been high enough for long enough such that the density above the
photosphere is ripe to produce SiO maser emission. As the mass-loss
continues at a gradually increasing rate, the critical densities for
the the H$_{2}$O and OH masers are reached at larger distances from
the star, and all three masers are switched on. At this point the star
may have shed enough of the H-rich envelope for the opacity in the
outer layers to drop, causing a rapid evolution back toward the blue.

Further empirical evidence for this evolutionary scenario is found in
RSGC1, in the form of a luminous yellow star, apparently similar in
nature to $\rho$~Cas \citep{RSGC1paper}. This star has a similar
luminosity to the brightest RSGs in the cluster (see
Fig.\ \ref{fig:maser}), consistent with the idea that the star has
evolved away from the tip of the RSG zone at constant bolometric
luminosity.

\subsection{The link between initial mass / post-SN remnant}
In RSGC1, there is the further possibility to make a direct link
between initial stellar mass, post-MS evolution and the type of object
left over after SN. This is due to the recent discovery of a TeV
gamma-ray source in the field of the cluster, which has subsequently
been confirmed from X-ray observations as a young pulsar
\citep{G-H08}. The H column density towards the pulsar is consistent
with the optical extinction of the cluster, strong evidence that the
two are physically associated. From the pulsar timing parameters,
Gotthelf \& Halpern determine that the pulsar is $\sim 10^{4}$yrs old,
meaning that the initial mass of the pulsar's progenitor must be
similar to that of the next most-massive star in the cluster which has
not yet gone SN. In RSGC1, the next most-massive stars are the RSGs
themselves, with \minit=18\msun \citep{RSGC1paper}. Hence, we can now
construct the evolutionary path for an 18\msun\ star:
\\ \\
MS - (BSG??) $\rightarrow$ RSG $\rightarrow$ YSG $\rightarrow$ (WN??)
$\rightarrow$ SN $\rightarrow$ neutron star
\\ \\
The evidence from RSGC1 that \minit=18\msun\ leave behind neutron
stars after SN is especially intriguing when discussed in the context
of the recent results from \citet{Smartt08}. These authors claim that
there is a statistically significant absence of Type-IIP SNe
progenitors with masses above 17\msun. As explanation for this result,
Smartt et al.\ suggest that stars with \minit\ slightly greater than
18\msun\ may retain enough of their initial mass by the end of their
lives to produce a black-hole, which then suppresses the brightness of
the SN explosion to below a detectable limit. The evidence from
RSGC1 that stars with \minit=18\msun\ produce neutron stars seems to
contradict this hypothesis, at least at Galactic metallicities.

\section{Summary and outlook}
In this review, I have discussed how young massive clusters, with
their large numbers of coeval massive stars, may be used to provide
direct links between stars of a certain initial mass, phases of
post-MS evolution, and even post-SN remnants. In doing so I discussed
case studies of three clusters in particular:

\begin{itemize}
\item {\bf The Arches Cluster -- } this cluster contains many stars
  with initial masses $\sim$100\msun, and so allows us to address the
  evolutionary status of the `WNLh' stars -- core H burning objects
  whose dense winds give them the appearance of WR stars. The H
  mass-fractions and luminosities of these stars appear to `trespass'
  into the domain of the LBVs, yet they do not seem to display the
  trademark variability. This raises the question of whether all stars
  above a certain mass limit do indeed pass through the LBV phase, or
  only those stars with certain initial conditions. 
\item {\bf Westerlund~1 -- } being the closest thing in our Galaxy to
  the Super Star Clusters seen in starburst galaxies such as M51 and
  M82, Wd~1 is a natural laboratory with which to test the simple
  stellar population models used to study unresolved clusters. The
  result that the binary fraction seems to be higher than 70\%, and
  that the WR population cannot be explained without a large degree of
  close binary interaction among the massive stars, serves as a note
  of caution when interpreting the integrated properties of unresolved
  systems.
\item {\bf RSGC1 -- } using this cluster we can make a link between
  maser emission and stars which are approaching the end of the RSG
  phase, as well as make a direct link between various evolutionary
  phases and post-SN remnants for `mid-range' massive stars with
  \minit=18\msun.
\end{itemize}

In addition to these objects, there are several more known young
clusters which may provide further evidence for the evolution of
massive stars through to SN. These include Cl~1806-20, which is known
to host an LBV and a highly magnetized neutron star
\citep{Figer05sgr,Bibby08}; Cl~1813-18, which again hosts a young
neutron star as well as WRs and an RSG \citep{Messineo08}; and
NGC3603, which hosts the most massive star known with a direct mass
measurement, 116$\pm$31\msun\ \citep{Schnurr08}.

%\acknowledgements %%% Text of acknowledgements runs on after this command.

%%% THE BIBLIOGRAPHY
%%%
%%% CONSULT SECTION 3 OF "INSTRUCTIONS FOR AUTHORS" FOR HOW TO USE NATBIB.
%%% AUTHORS ARE ENCOURAGED TO USE EITHER THE "THEBIBLIOGRAPY" ENVIRONMENT
%%% BY UNCOMMENTING (DELETING THE "%" SYMBOL) THE COMMANDS BELOW, OR BY
%%% USING THE BIBTEX ENVIRONMENT. TO FIND OUT WHICH IS APPLICABLE TO YOUR
%%% CONTRIBUTION, CONSULT THE VOLUME EDITORS FOR YOUR PROCEEDINGS.
%%%

%\begin{thebibliography}{}
%\bibitem[]{}
%\bibitem[]{}
%\bibitem[]{}
%\bibitem[]{}
%\bibitem[]{}
%\bibitem[]{}
%\bibitem[]{}
%\bibitem[]{}
%\bibitem[]{}
%\bibitem[]{}
%\bibitem[]{}
%\bibitem[]{}
%\end{thebibliography}

%\bibliographystyle{/fat/Data/bibtex/apj}
%\bibliography{/fat/Data/bibtex/biblio}

\end{document}